\documentclass[runningheads]{llncs}
\usepackage{graphicx}
\usepackage[backend=biber, style=alphabetic, sorting=ynt]{biblatex}
\usepackage{hyperref}
\usepackage{graphicx}
\hypersetup{
    colorlinks=true,
    linkcolor=blue,
    filecolor=magenta,      
    urlcolor=cyan,
    pdftitle={Overleaf Example},
    pdfpagemode=FullScreen,
    }

\begin{document}
 \title{Whole-body tumor segmentation of ${}^{18}F$-FDG PET/CT using a cascaded and ensembled convolutional neural networks}
\author{Ludovic Sibille\inst{1}  \and
Xinrui Zhan\inst{1} \and
Lei Xiang\inst{1}}
\authorrunning{L. Sibille et al.}
\institute{Subtle Medical, USA 
%\and
%Springer Heidelberg, Tiergartenstr. 17, 69121 Heidelberg, Germany
%\url{http://www.springer.com/gp/computer-science/lncs} 
%\and
}
\maketitle              % typeset the header of the contribution
\begin{abstract}
~\\
\textbf{Background}: A crucial initial processing step for quantitative PET/CT analysis is segmentation of tumor lesions enabling accurate feature extraction, tumor characterization, oncologic staging and image-based therapy response assessment. Manual lesion segmentation is however associated with enormous effort and cost and is thus infeasible in clinical routine. \\
\textbf{Goal}: The goal of this study was to report the performance of a deep neural network designed to automatically segment regions suspected for cancer in wholebody 18F-FDG PET/CT images in the context of the AutoPET challenge.\\
\textbf{Method}: A cascaded approach was developed where a stacked ensemble of 3D UNET CNN processed the PET/CT images at a fixed 6mm resolution. A refiner network composed of residual layers enhanced the 6mm segmentation mask to the original resolution.\\
\textbf{Results}: 930 cases were used to train the model. 50\% were histologically proven cancer patient and 50\% were healthy controls. We obtained a dice=0.68 on 84 stratified test cases. Manual and automatic Metabolic Tumor Volume (MTV) were highly correlated ($R^2=0.969$, $Slope=0.947$). Inference time was 89.7 seconds on average.\\ \textbf{Conclusion}: The proposed algorithm accurately segmented regions suspicious for cancer in wholebody ${}^{18}F$-FDG PET/CT images.

\end{abstract}

\section{Introduction}
PET with fluorine 18 (18F) fluorodeoxyglucose (FDG) has a substantial impact on the diagnosis and clinical decisions of oncological diseases. 18F-FDG uptake highlights regions of high glucose metabolism that include both pathological and physiological processes. 18F-FDG-PET/CT adds value to the initial diagnosis, detection of recurrent tumor, and evaluation of response to therapy in lung cancer, lymphoma and melanoma.

FDG PET images are interpreted by experienced nuclear medicine readers that identify foci positive for FDG uptake that are suspicious for tumor. This classification of FDG positive foci is based on a qualitative analysis of the images and it is particularly challenging for malignant tumors with a low avidity, unusual tumor sites, with motion or attenuation artifacts, and the wide range of FDG uptake related to inflammation, infection, or physiologic glucose consumption \cite{doi:10.1080/08998280.2005.11928089, PETFP}.

A crucial initial processing step for quantitative PET/CT analysis is segmentation of tumor lesions enabling accurate feature extraction, tumor characterization, oncologic staging and image-based therapy response assessment. Manual lesion segmentation is however associated with enormous effort and cost and is thus infeasible in clinical routine. 

This paper summarizes the approach used in the AutoPET challenge \cite{AutoPET} to segment suspicious lesions in wholebody FDG-PET/CT images.

\section{Material and methods}
\subsection{Patients}

The challenge cohort consists of patients with histologically proven malignant melanoma, lymphoma or lung cancer who were examined by FDG-PET/CT in two large medical centers (University Hospital Tübingen, Germany and University Hospital of the LMU in Munich, Germany). Two expert radiologists with 5 and 10 years of experience annotated the dataset using manual free-hand segmentation of identified lesions in axial slices.

\begin{figure}
\hfill
\includegraphics[height=5.5cm]{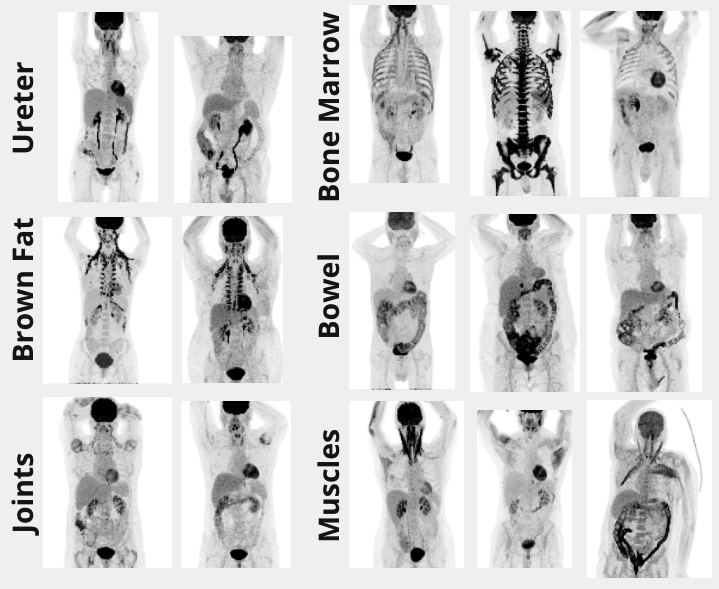}
\hfill
\includegraphics[height=5.5cm]{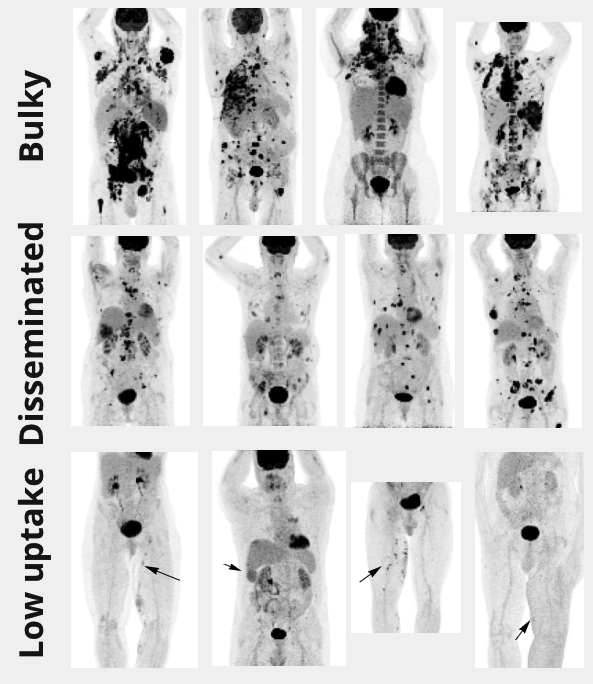}
\hfill
\caption{Patients illustrating the large variations in appearance of normal uptake patterns (left) and their within class variations. On the right are showed FDG uptake variations of cancerous lesions.}
\label{fig:overview}
\end{figure}

In total, 900 patients were acquired in 1014 different studies. 50\% of the patients were negative control patients. Refer to \cite{AutoPET} for additional information on the acquisition protocol. As can be seen in figure \ref{fig:overview}, a wide range of normal FDG uptakes were present such as brown fat, bone marrow, bowel, muscles, ureter and joints as well as within class variations. Similarly, patients with lesions showed large variations such as bulky, disseminated or low uptake patterns.

The dataset was split in 2 independent splits (development, test) at the patient level to avoid data leakage. The development dataset was further split in 15-fold cross-validation sets. Given the long tail distributions of the lesions, all the splits were stratified by overall lesion volume and number of lesions to minimize the model variance trained on different data slits.

\subsection{Data pre-processing}
The data pre-processing aimed to mimic the ranges a radiologist would use to discriminate the various FDG uptakes and help the neural network disentangle the inputs. The PET SUV and CT volumes were processed as followed:
\begin{itemize}
    \item SUV: The PET SUV image was mapped from (0, 30) SUV to (0, 1) range. This captured most of the PET intensities
    \item CT: The PET matched CT image mapped from (-150, 300) to (0, 1). This captured the most important patterns of the CT
    \item CT\_Soft: We used the range (-100, 100) to focus on the soft-tissue intensities 
    \item CT\_Lung: Same logic with CT\_Lung. Instead, the value ranged (-1000, -200) to capture the intensities of the lung tissues
    \item SUV\_hot: range is (2, 10) of the PET SUV to focus on the mid-range intensities of the lesions. This was useful for lesions with low uptake
\end{itemize}

\subsection{Data augmentation}
We evaluated various data augmentation schemes and only the following ones lead to marked improvements on the validation splits:
\begin{itemize}
    \item Random axis flip for all three dimensions
    \item Random affine transformation which included random rotations and isotropic scaling
    \item PET only augmentation: random Gaussian blur, brightness, contrast and gamma transforms
\end{itemize}
The refiner network had an additional transformation that resampled the data using random spacing to make the refiner completely independent from the original input resolution. 

\subsection{Model}
\begin{figure}[h]
\includegraphics[width=\textwidth]{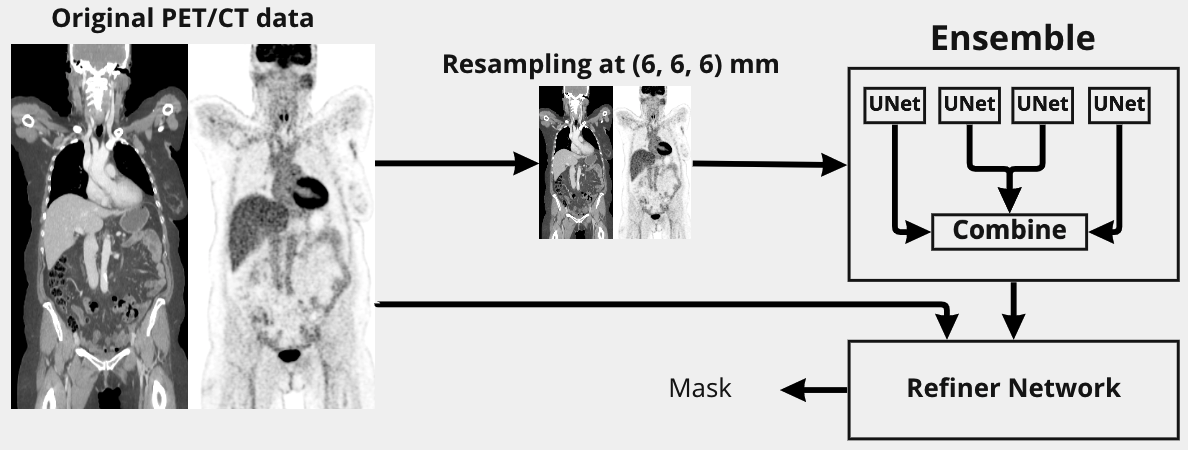}
\caption{Overview of the proposed approach: the original data is downsampled to an isometric 6mm resolution. An ensemble of 4 customized UNET models are linearly aggregated to output a low resolution segmentation mask. The original PET/CT images and the 6mm segmentation mask are then sent to a refiner network to recover the original resolution.}
\label{fig:overview_model}
\end{figure}

At a high level, the proposed model approach can be decomposed in 2 modules: the first module had a large field of view to analyse at a coarse level global patterns and long range dependencies. The analyzed images were fixed in resolution and downsampled to 6mm to make the module resolution independent. The goal of the second module was to refine the coarse segmentation found by the first module using the original image (see figure \ref{fig:overview_model})

The first module was composed of an ensemble of UNET like neural networks. UNET was modified to take as input 5x128x96x96 images. The channels were set to [64, 96, 128, 156] with a stride of 2 for each layer. Besides the change of channels, we also modified the middle block to have large kernels $9^3$ to encourage the detection of long range dependencies. Leaky ReLU was used as activation function and instance normalization as normalization layer. We employed deep supervision \cite{DeepSupervision} to train the UNET. 4 UNETs were trained on different splits of the development dataset and their outputs were linearly weighted \cite{Breiman2004StackedR} to form the 6mm intermediate segmentation mask. 

The refiner network took as input 5x64x64x64 images at original resolutions with the 6mm segmentation mask resampled to match the original image resolution. It was composed of a stem block with $9^3$ kernel followed by 4 x ($3^3$ convolution, leaky ReLU, instance norm) residual blocks with a final $3^3$ convolution to calculate the segmentation mask.

The models were trained to minimize the following loss:
$$
loss = dice\_loss + 0.5 * cross\_entropy\_loss + 2 * sensitivity\_loss
$$

Each model (UNET, ensemble stacking, refiner) were trained separately. We used AdamW \cite{loshchilov2018fixing} as optmizer with a learning rate of $1e-3$, weight decay of $1e-6$ and a decayed cosine warm restart scheduler (T=200 epochs, decay=0.9 for each period). Finally, gradient clipping was applied to stabilize the training.

We report dice, dice of cases with foreground segmentation (`foreground dice`), sensitivity, False positive (FP) volume (volume of incorrectly segmented uptake that did not overlap with true segmentation), False negative (FN) volume (volume of incorrectly segmented uptake that did not overlap with true segmentation), metabolic tumor volume (MTV) for the evaluation of our method.

\subsection{Post-processing}
We tried to add sequence based models (RNN, LSTM \& GRU) to reduce the false positive segmentations. The segmentation map outputed by the refiner network was relabelled using connected components. All features of the penultimate layer of the segmentation model belonging to the same connected components were averaged. In addition, high-order features were added (e.g., tumor volumes, SUV max, SUV std, position in volume, shape descriptors). Finally, a sequence model was used to re-classify the connected components. Intuitively, leveraging the segmentation as sequences may be easier to model long range dependencies and high order features may mimic features a radiologist may use. Unfortunately, results were very similar and for simplicity sake's, we removed it from our processing pipeline. 

\subsection{Implementation}
The method was developed using Python 3.9 \cite{10.5555/1593511} and Pytorch 1.11 \cite{NEURIPS2019_9015}. Scripts for training and deployment are publicly available here:\\
\url{https://github.com/civodlu/AutoPET2022}

\section{Results}
\begin{figure}[h]
\includegraphics[width=\textwidth]{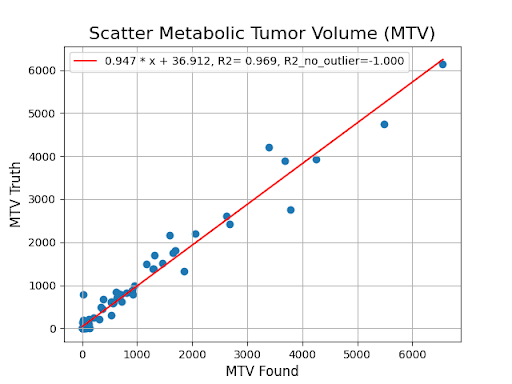}
\caption{Scatter plot of the manually and automatically segmented metabolic tumor volumes on the test data. Automatic and manual methods were found to be equivalent.}
\label{fig:scatter_mtv}
\end{figure}

930 cases were used for the training of the models and 84 cases were held out for the final evaluation. Qualitatively, the model produced accurate segmentations. It understood the typical variations in appearance of normal uptakes such as hot organs (e.g., kidneys, bladder, brain, heart, ureter), brown fat, bone marrow, inflammation (e.g., bowel and joints), muscles. Similarly, the segmentations of lesions was accurate, all large lesions were segmented (see Figure \ref{fig:scatter_mtv}) and Table \ref{table:results_test}. Segmenting very small lesions or lesions with low FDG avidity was challenging.

For the various trained models, we found that combining a dice loss with cross-entropy loss lead to a more stable training of the models. Similarly, deep supervision and gradient clipping lead to improved training stability. The sensitivity loss encouraged the models to segment the smaller lesions but at the cost of additional false positives.

The whole pipeline took on average 89.7 seconds on a single V100 GPU per case, although no attempt were made to minimize inference speed.

Automatic and manually segmented metabolically tumor volumes were equivalent as shown in Figure \ref{fig:scatter_mtv}

\begin{table}[h]
\begin{tabular}{ |c|c|c|c|c|c|c|c|c| }
\hline
 Config & Dice &  Dice Foreground & FN & FP & Sensitivity & MTV found & MTV & time (s) \\
\hline
 M0 & 0.4819 & 0.6643 & \textbf{15.39} & 15.53 & \textbf{0.7426} & 558.6 & 485.4 & 4.94 \\
\hline
 M1 & 0.4776 & 0.6583 & 21.63 & 15.52 & 0.7097 & 530.6 & 485.4 & 4.25 \\
\hline
 M2 & 0.4709 & 0.6618 & 22.07 & 26.30 & 0.6939 & 525.0 & 485.4 & 4.06 \\
\hline
 M3 & 0.5028 & 0.6658 & 26.33 & 12.51 & 0.7072 & 544.5 & 485.4 & 4.09 \\
\hline
\hline
 Ensemble & \textbf{0.5049} & \textbf{0.6823} & 19.59 & \textbf{11.59} & 0.7255 & 536.3 & 485.4 & 8.91 \\
\hline
\hline
 Full &  0.4942 & 0.6813 & 16.46 & 17.85 & 0.6446 & 413.2 & 485.4 & 89.7 \\
\hline
\end{tabular}
\caption{Results of the models on the held out test split. `Dice Foreground` represents the dice of the cases that have a segmentation. `Dice` are the dice results on all the cases. Note that in the special case where a segmentation is found on a patient that has no segmentation, a dice of `0` is attributed to this case, which this is different from the challenge's dice calculation. If no segmentation is found on a case without segmentation, this case is discarded for the dice calculation (this differs again from the challenge which attributed `1` in this case explaining the differences in the leaderboard). `FN` is the false negative volumes, `FP` is the false positive volume, `MTV` represents the average volume of the segmentations expressed in voxels.}\label{table:results_test}
%\end{center}
\end{table}

Table \ref{table:results_test} shows that resulting dice were very similar for the models (M0-M3) trained on different splits. However, there were large variations in terms of False Negative (FN) and False Positive (FP) volumes as well as sensitivity despite an identical processing pipeline. The ensemble network had an improved `dice` and `dice foreground`. Finally, the refiner network was able to keep very similar performance characteristics while operating on the full scale image with FP and FN volumes adequately rebalanced.

\clearpage
\section{Conclusion}
The proposed algorithm accurately segmented regions suspicious for cancer in wholebody FDG-PET/CT images.

\printbibliography

\clearpage

\section{Appendix}

\begin{figure}[h]
\includegraphics[width=0.9 \textwidth]{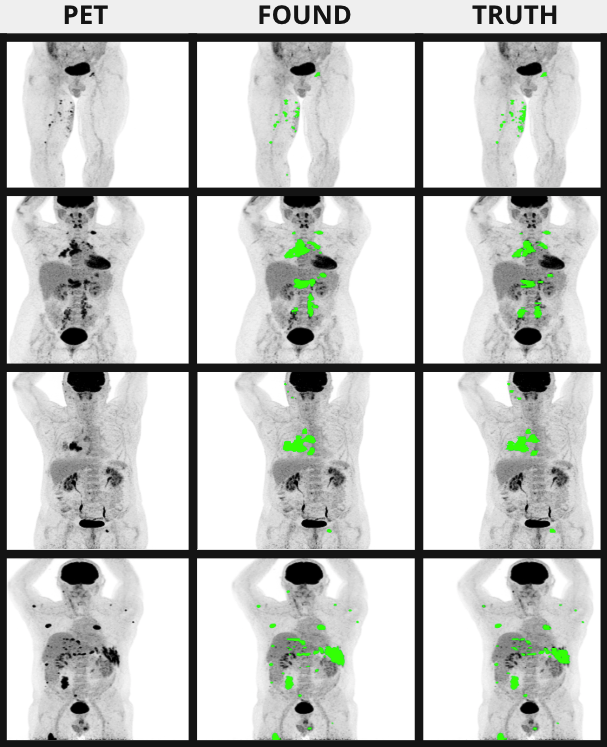}
\caption{Example of typical automatic segmentation results on the test data.}
\label{fig:average_examples}
\end{figure}

\begin{figure}[h]
\includegraphics[width=0.9 \textwidth]{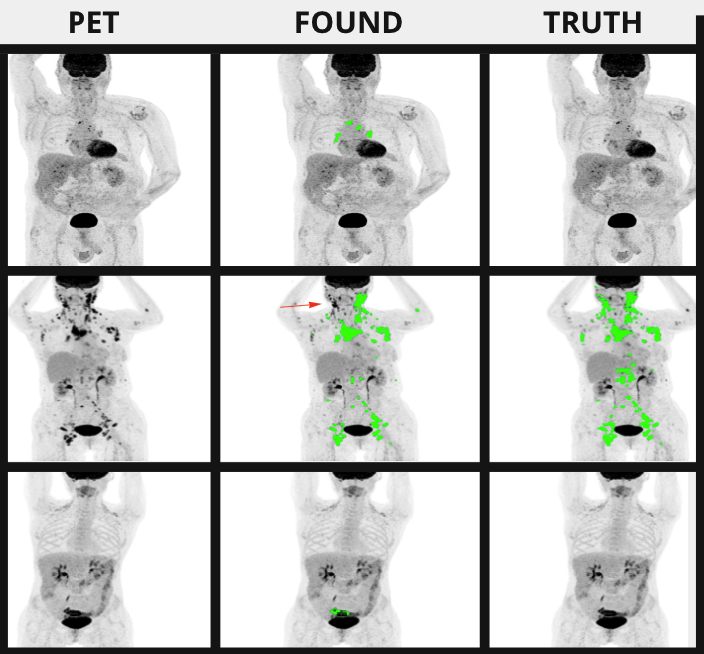}
\caption{Example of failure cases: first row, infectious disease are mistaken for lesions probably due to the paucity of the training data. Second row: right cervical lymph nodes are missed. Third row: inflammation of sigmoid colon is mistaken for a lesion.}
\label{fig:average_examples}
\end{figure}

\end{document}